\documentclass[final,5p,times,twocolumn,10pt]{elsarticle}

\usepackage{amsmath,lscape,epsfig}
\usepackage{amsfonts}
\usepackage{amsmath}
\usepackage{amssymb}
\usepackage{slashed}
\usepackage[utf8]{inputenc}
\usepackage{graphicx,color}
\usepackage{epstopdf}
\usepackage{float}
\usepackage{hyperref}

\def\beq{\begin{equation}}
\def\eeq{\end{equation}}
\def\beqa{\begin{eqnarray}}
\def\eeqa{\end{eqnarray}}
\def\ban{\begin{eqnarray*}}
\def\ean{\end{eqnarray*}}
\def\bi{\begin{itemize}}
\def\ei{\end{itemize}}

\journal{Physics Letters B}

\begin{document}

\begin{frontmatter}

\title{$\pi_0$ pole mass calculation in a strong magnetic field  and lattice constraints}

\author{Sidney S. Avancini\corref{mycorrespondingauthor}}
\cortext[mycorrespondingauthor]{Corresponding author}
\ead{sidney.avancini@ufsc.br}

\address{Departamento de F\'{\i}sica, Universidade Federal de Santa
  Catarina, 88040-900 Florian\'{o}polis, Santa Catarina, Brazil }

\author{Ricardo L. S. Farias}
\ead{rfarias@kent.edu}
\address{Departamento de F\'{\i}sica, Universidade Federal de Santa Maria, 
97105-900 Santa Maria, RS, Brazil \\
Physics Department, Kent State University, Kent, OH 44242, USA }

\author{Marcus Benghi Pinto}
\ead{marcus.benghi@ufsc.br}
\address{Departamento de F\'{\i}sica, Universidade Federal de Santa
  Catarina, 88040-900 Florian\'{o}polis, Santa Catarina, Brazil }

\author{William R. Tavares}
\ead{william.tavares@posgrad.ufsc.br}
\address{Departamento de F\'{\i}sica, Universidade Federal de Santa
  Catarina, 88040-900 Florian\'{o}polis, Santa Catarina, Brazil }
  
\author{Varese S. Tim\'oteo}
\ead{varese@ft.unicamp.br}
\address{Grupo de \'Optica e Modelagem Num\'erica (GOMNI), Faculdade de Tecnologia, 
Universidade Estadual de Campinas - UNICAMP, Brazil }

\begin{abstract}
The $\pi_0$ neutral meson pole mass is calculated in a strongly magnetized medium using the 
SU(2) Nambu-Jona-Lasinio model within the random phase approximation (RPA) at zero temperature 
and zero baryonic density. We employ a magnetic field dependent coupling, $G(eB)$, fitted
to reproduce  lattice QCD results for the quark condensates. Divergent quantities are handled with a magnetic 
field independent regularization scheme in order to avoid unphysical oscillations. A comparison 
between the running and the fixed couplings reveals that the former produces results much closer to the 
predictions from recent lattice calculations. In particular, we find that the $\pi_0$ meson mass systematically 
decreases when  the magnetic field increases while the scalar mass remains almost constant. 
We also investigate how the magnetic background influences other mesonic properties such as  
$f_{{\pi}_0}$ and $g_{\pi_0 q q}$.
\end{abstract}

\end{frontmatter}


\section{Introduction}
Very recently \cite{nosso1} the ${\pi}_0$ pole mass was calculated in a strongly magnetized medium using 
the two flavor Nambu-Jona-Lasinio model \cite{NJL-122,lemmer,kleva,hatku,buballa}. With this aim, a 
generalization of the standard $B=0$ evaluation, within the RPA (or ladder approximation) \cite{kleva}, 
has been carried out to take into account the presence of a  strong magnetic field. 
The formalism employed in Ref.\cite{nosso1} is based on the use of field dependent Feynman propagators 
\cite{gus,kuz} and a field independent  regularization scheme (MFIR) \cite{klimenko,nosso2} which separates 
divergent vacuum contributions from finite thermo-magnetic ones. The advantages of using the MFIR, 
specially in connection with high density effects, have been recently  discussed in great 
detail \cite{norberto,ricardo}. 
In those references it has been discussed that calculations which employ $B-$dependent regularizations display, 
in general, a spurious behavior specially at high magnetic field values which constitute a serious
drawback for their 
use in many physical situations of interest as discussed in the literature 
\cite{mandal,garb-1,garb-2}. Other powerful tools which have been employed in the evaluation of pionic observables include 
chiral perturbation theory (ChPT)~\cite{andersen,agasian} and effective quark-antiquark lagrangians~\cite{simonov1,simonov2}.
 
The magnetic catalysis phenomenon (MC), ie., the enhancement of the quark condensate when the magnetic 
field increases is a common characteristic of all mean field calculations. However, accurate lattice calculations 
at zero chemical potential and finite temperature predict exactly the opposite behavior close to the (pseudo) 
critical temperature. This effect has been called inverse magnetic catalysis (IMC) or magnetic inhibition 
\cite{fukushima,review1,review2}.  \footnote{ An alternative definition of IMC at finite chemical potential and 
moderate magnetic field can be found in the literature\cite{andreas}. }

In  some recent calculations \cite{ric1,ric2} thermo-magnetic effects were included in the standard two-flavor 
version of the NJL model, where the coupling constant $G$ has been allowed to become temperature and 
magnetic field dependent,i.e., $G \to G(eB,T)$, and fitted according to recent lattice results for the quark 
condensates, 
emulating the running of the QCD coupling constant with magnetic field and temperature, and thus incorporating IMC.  
It has been shown in Ref. \cite {ric2}  that the thermodynamic quantities calculated with the SU(2) NJL model using 
$G(eB,T)$ behave in accordance with   lattice predictions, something which is impossible to obtain in 
the traditional
calculations with constant coupling.
 
In the present letter we improve the application of Ref. \cite {nosso1} by determining an accurate running 
coupling, $G(eB)$, at vanishing temperatures, following the procedure  introduced recently in Ref.~\cite {ric2} where the 
thermo-magnetic running (in the high-$T$ limit) was determined. We show that in this case 
the neutral pion mass, 
which represents the soft mode, is in excellent numerical agreement with  lattice QCD 
simulations  ~\cite {endrodidata1,endrodidata2}. 
At the same time we find that the scalar meson mass remains almost constant for a wide range of $eB$ values. 
This interesting result, which is a byproduct of the  stability of the effective quark mass within our approach, 
should be contrasted to the linear increase found when a fixed coupling is used, causing the scalar mode to decouple
at strong magnetic fields. 

We also investigate the neutral pion decay constant and predict that this quantity increases with $B$ in a way 
compatible with the Gell-Mann-Oakes-Renner relation. Finally, we also predict that the meson-quark couplings 
decrease with increasing magnetic fields. The Letter is organized as follows. In the next section
we present the  model and the  formalism. The numerical 
results are discussed in Sec. 3 and our concluding remarks are presented in Sec. 4.

\section{General formalism}

Let us start by reviewing the main steps related to the evaluation of the mesonic properties using  the RPA 
formalism within the MFIR framework as done in Ref. \cite {nosso1}. We also present the ansatz for the magnetic 
coupling at vanishing temperatures.

\subsection{Meson properties under strong magnetic field}

In the presence of a magnetic field the standard two-flavor NJL model  is described by
\begin{eqnarray}
\mathcal{L}&=&\bar{\psi}_f\left(i \slashed D - \tilde{m}\right)\psi_f
+G\left[(\bar{\psi}_f\psi_f)^{2}+(\bar{\psi}_f i\gamma_{5}\vec{\tau}\psi_f)^{2}\right]\nonumber \\
&-&\frac{1}{4}F^{\mu\nu}F_{\mu\nu} ~,
\end{eqnarray}
where a sum over repeated $f$ is implied. The electromagnetic gauge field is represented by  $A^\mu$, 
$F^{\mu\nu} = \partial^\mu A^\nu - \partial^\nu A^\mu$ , $\vec{\tau}$ is the isospin matrix, 
the coupling constant by $G$ while $Q$=diag($q_u$= $2 e/3$, $q_d$=-$e/3$) represents the charge matrix,
$D^\mu =(i\partial^{\mu}-QA^{\mu})$ is the covariant derivative, ${\psi}^{\rm T}$=$(\psi_u,\psi_d)$  is the 
quark fermion field and $\tilde{m}$ = diag($m_u,m_d$)  represents the bare quark mass matrix.
 
Here, we  adopt the Landau gauge, i. e., $A^{\mu}=\delta_{\mu 2}x_{1}B$, thus $\vec{B}=B{\hat{e_{3}}}$. 
Then, in the mean field approximation the NJL lagrangian is given by \cite{buballa}
\begin{equation}
 \mathcal{L}=\bar{\psi}_f\left(i\slashed D-M_f \right)\psi_f+G \left \langle \bar{\psi}_f \psi_f \right \rangle^{2}-
 \frac{1}{4}F^{\mu\nu}F_{\mu\nu}~,
\end{equation}
\noindent
where $\langle \bar{\psi}_f \psi_f \rangle$ represents the quark condensates. The effective quark mass for a given flavor is

\begin{equation}
 M_i=m_i - 2G [\left \langle \bar{\psi}_i\psi_i \right \rangle +  \langle \bar{\psi}_j\psi_j  \rangle] \;,
\label{gap}
\end{equation}
with $i,j=u,d$ and $i \ne j$.
 Note that by taking $m=m_u$=$m_d$, 
as we do here, one may set $M_u=M_d=M$ since the different condensates enter in a symmetric manner. 
It has been shown in Ref.\cite{nosso1}  that in the RPA approximation  the $\pi_0$ meson mass in a magnetized 
medium can be calculated selecting the quantum numbers associated to the neutral pion. From the Bethe-Salpeter 
equation one obtains:
\begin{equation}
  (ig_{{\pi_0} q q})^2~iD_{\pi_0}(k^2)=\frac{2iG}{1-2G\Pi_{\rm PS}(k^{2})} ~, \label{BS}
\end{equation}
As usual in the last equation  the left hand side of the equality is calculated by representing  the quark-pion 
interaction with the following Lagrangian density\cite{kleva}:
\begin{equation}
 \mathcal{L}_{\pi q q}=i g_{\pi qq} \bar{\psi} \gamma_5 \vec{\tau}\cdot \vec{\pi} \psi~,
\end{equation}
\noindent where  $\vec{\pi}$ stands for the pion field while $g_{\pi q q}$ represents the coupling strength
between pions and quarks. Both sides of eq.(\ref{BS}) can be calculated  using the standard meson 
propagator \cite{ryder},  
\begin{equation}
  D_{\pi_0}(k^2) = \frac{1}{k^2-m_{\pi_0}^2} ~, \label{meson}
\end{equation}
as well as the quark (dressed) propagator in a magnetic medium\cite{gus,kuz},
\begin{equation}
 S_{q}(x,x')=e^{i\Phi_q (x,x')}\sum_{n=0}^{\infty}{S}_{q,n}(x-x')~,~q=u,d~. \label{prop1}
\end{equation}
\noindent  
The quark propagator in a strong magnetic field is given by the product of a gauge dependent 
factor, $\Phi_q(x,x')$, called Schwinger phase, times a translational invariant term and its explicit expression 
can be found in Ref. \cite{kuz}. In the present calculation, which involves only neutral particles, 
the Schwinger phase 
cancels out. Through the use of standard Feynman rules the pseudo-scalar  polarization loop reads 
(see Ref.\cite{nosso1} for further technical details):
\begin{eqnarray}
\frac{1}{i}\! \Pi_{\rm PS}(k^{2})&=&\!-\!\sum_{q=u,d}\int\frac{d^{4}p}{(2\pi)^{4}}
Tr\left[i\gamma_{5}iS_q \left(p+\frac{k}{2}\right)
i\gamma_{5}\right.\nonumber\\
&\times&\left. i S_q \left(p-\frac{k}{2}\right)\right]~.  \label{loop-ps}
\end{eqnarray}
As shown in Ref. \cite {nosso1} an analogous expression can be obtained for the scalar channel. 
Then, from Eq.(\ref{BS}), one can obtain the $\pi_0$ mass pole as:
\begin{equation}
 1-2G~\Pi_{\rm PS}(k^{2})|_{k^2=m^2_{\pi_0}} = 0~. 
 \label{polemass}
\end{equation}
The explicit expression 
for the pseudoscalar polarization loop, Eq.(\ref{loop-ps}), is given by\cite{nosso1}:
\begin{multline}
 \frac{1}{i} \Pi_{\rm PS}(k^{2}_{\parallel})= -i\left(\frac{M-m}{2MG}\right) \\ 
- \sum_{q=u,d} \beta_{q} 
 N_{c}\frac{k_{\parallel}^{2}}{(2\pi)^{3}}\sum_{n=0}^{\infty}g_{n} I_{q,n}(k_{\parallel}^{2})~, \label{polar}
\end{multline} 
\noindent where 
\begin{equation}
 I_{q,n}(k_\parallel^2)=\!\int\! d^{2}p_{\parallel}\frac{1}{[p^{2}_{\parallel}-M^{2}-2\beta_{q} n]
 [(p+k)_{\parallel}^{2}-M^{2}-2\beta_{q} n]}. 
 \label{Iqn0}
\end{equation}
where $\beta_q$=$|q_q|B$, $q$=(u,d), $N_c$=3, $g_{n}=2-\delta_{n0}$ , $p_{\parallel}=p_{0}-p_{3}$, and 
$k_{\parallel}=k_{0}-k_{3}$.
Therefore, from Eq.(\ref{polemass}), the $\pi_0$ mass can be written as:
 \begin{equation}
 m_{{\pi}_0}^{2}(B)=-\frac{m}{M(B)}
 \frac{1}{ 4iG N_{c}N_{f} I(m_{\pi_0}^2,B) }~, \label{mpion}
\end{equation}
where
\begin{equation}
 I(m_{\pi_0}^2,B) = \frac{1}{4(2\pi)^{3}}
 \displaystyle{\sum_{q=u,d} \beta_{q} \sum_{n=0}^{\infty}
 g_{n} I_{q,n}(k_\parallel^{2}=m_{\pi_0}^2) }~. \label{mpion_int}
\end{equation}
The $\sigma$-meson mass, $m_\sigma$, is readily evaluated in a completely analogous fashion 
by calculating the scalar polarization loop. This procedure yields \cite{nosso1} :  
\begin{equation}
m_{\sigma}^{2}(B)=4M^{2}(B)+m_{\pi_{0}}^{2}(B)\,.
\end{equation}
Next, the pion decay constant is given by the expression:
\begin{equation}
 f_{\pi_{0}}^{2}(B)=-i\sum_{u,d}\frac{\beta_{q}}{(2\pi)^{3}}N_{c}M^{2}
 \sum_{n=0}^{\infty} g_{n}I_{q,n}(0) ~, \label{fpion}
\end{equation}
\noindent where $I_{q,n}(0) \approx I_{q,n}({{m_\pi}_0}^{2})$. The following
identity can be obtained from Eqs.~(\ref{mpion},\ref{fpion}) 
\begin{equation}
 m_{\pi_0}^2(B)f_{\pi_{0}}^{2}(B)= \frac{m~M(B)}{2G}~. 
 \label{mmpi}
\end{equation}
In the next section we perform an explicit numerical analysis concluding that the approximation 
$I_{q,n}(0) \approx I_{q,n}({m_{\pi_0}}^{2})$ provides results that differ from the exact one 
only by about $1\%$ or less.   

The gap equation, Eq.~(\ref{gap}), can be used in order to eliminate the coupling constant $G$
so that the Gell-Mann-Oakes-Renner (GOR) relation in a magnetic medium is recovered.

\begin{equation}
 m_{\pi_{0}}^{2}(B) f_{\pi_{0}}^{2}(B)= - m \left \langle \overline{\psi}_f\psi_f\right\rangle(B).
 \label{gor}
\end{equation}

In Ref.\cite{nosso1} the loop integral Eq.(\ref{Iqn0}) was obtained as
\begin{equation}
 I({k_\parallel}^2,B)=I_{vac}(k_\parallel^2)+I(k_\parallel^2,B) ~,  \label{pol_int}
\end{equation}
where
\begin{equation}
 I_{vac}({k_\parallel}^2) = \frac{i}{8 \pi^2} \int_{0}^{1} dx \left[    
 \sinh^{-1} \left(\frac{\Lambda}{\overline{M}} \right) - 
 \frac{\Lambda}{\sqrt{\Lambda^2+\overline{M}^2} }  \right] \nonumber ~, \\
 ~\label{ivac3} 
\end{equation}
and
\begin{eqnarray}
 I(k_\parallel^2,B)&=&\frac{i\pi}{4(2\pi)^3}\sum_{q=u,d}
 \int^{1}_{0}dx  \left[ -\psi\left(x_q+1 \right)  +
 \frac{1}{2x_q}\right.\nonumber\\
 &+&\left.  \ln x_q  \right]  ~, \label{int00}
\end{eqnarray}
with 
\begin{equation}
 x_q = \frac{\overline{M}^{2}(k_\parallel^2)}{2\beta_{q}} ~~,~~
 \overline{M}^{2}(k_{\parallel}^2)=M^{2}-x(1-x)(k_{\parallel}^{2}).
\end{equation}
Following the MFIR prescription \cite{nosso2}, we have disentangled overlapping divergences by dividing 
the polarization integral, Eq. (\ref{pol_int}), into two terms: the first takes into account divergent vacuum 
fluctuations and can be  regularized through a non-covariant three-momentum cutoff, while the second, 
Eq. (\ref{int00}), represents the finite contribution due to magnetized medium. Note that using the MFIR 
scheme one recovers the usual vacuum term. 
\begin{figure}[h]
\centering
\includegraphics[width=8.5cm]{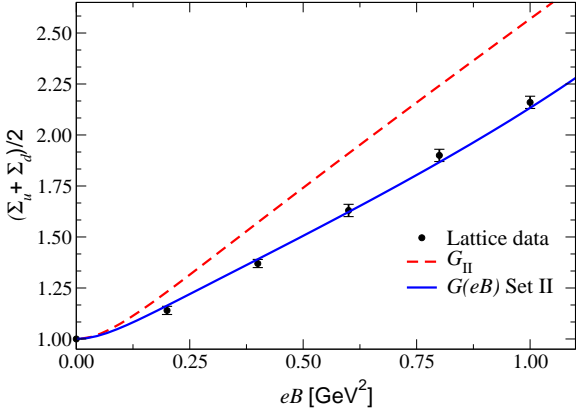}\\
\includegraphics[width=8.5cm]{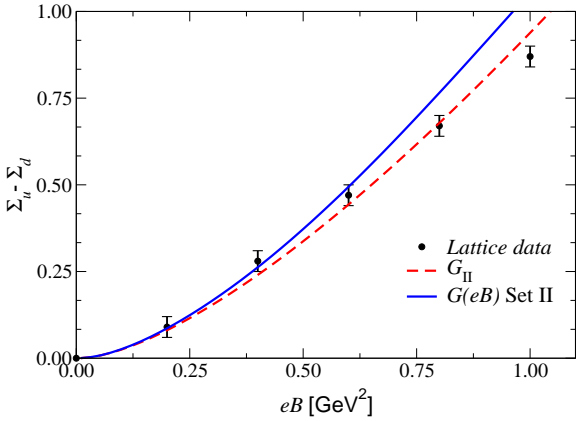}\\
\caption{Condensates average and difference as functions of $eB$ for the NJL model with 
$G_{II}$, $G(eB)$ compared to lattice QCD calculations from Ref.~\cite{fodor}.}
\label{fig1}
\end{figure}

\begin{figure}[h]
\centering
\includegraphics[width=8.5cm]{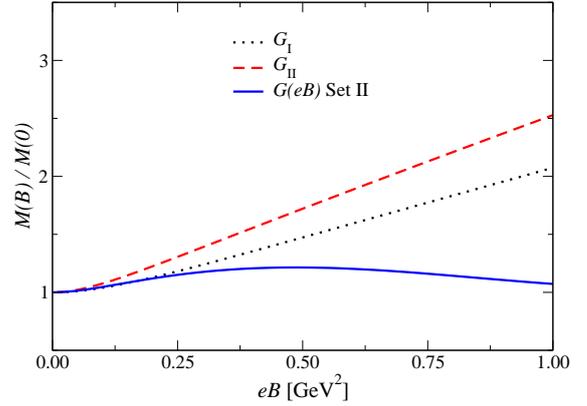}\\
\caption{Normalized constituent quark mass as a function of $eB$ for the NJL model with different coupling schemes.}
\label{fig2}
\end{figure}
\begin{figure}[h]
\centering
\includegraphics[width=8.5cm]{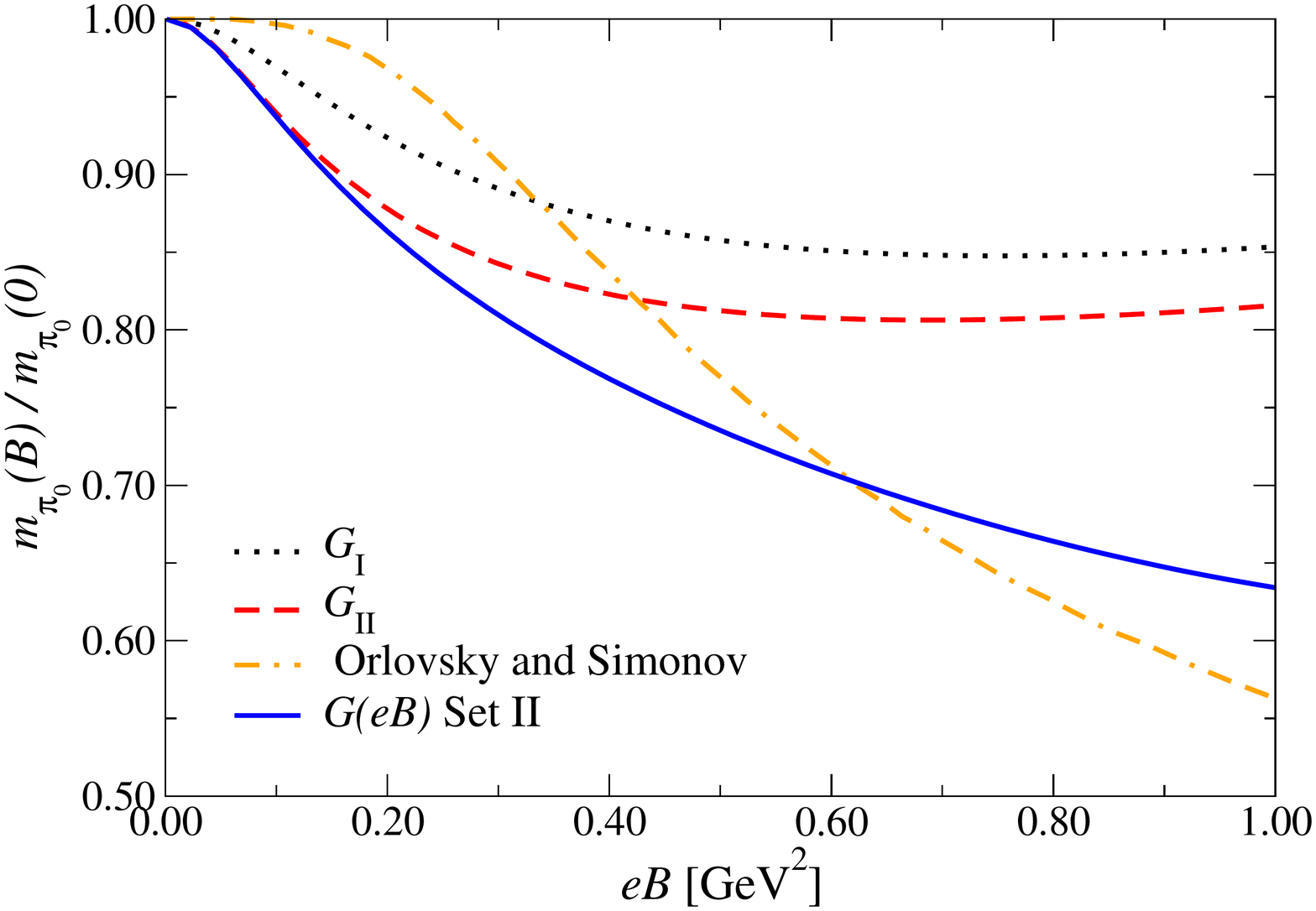}
\includegraphics[width=8.5cm]{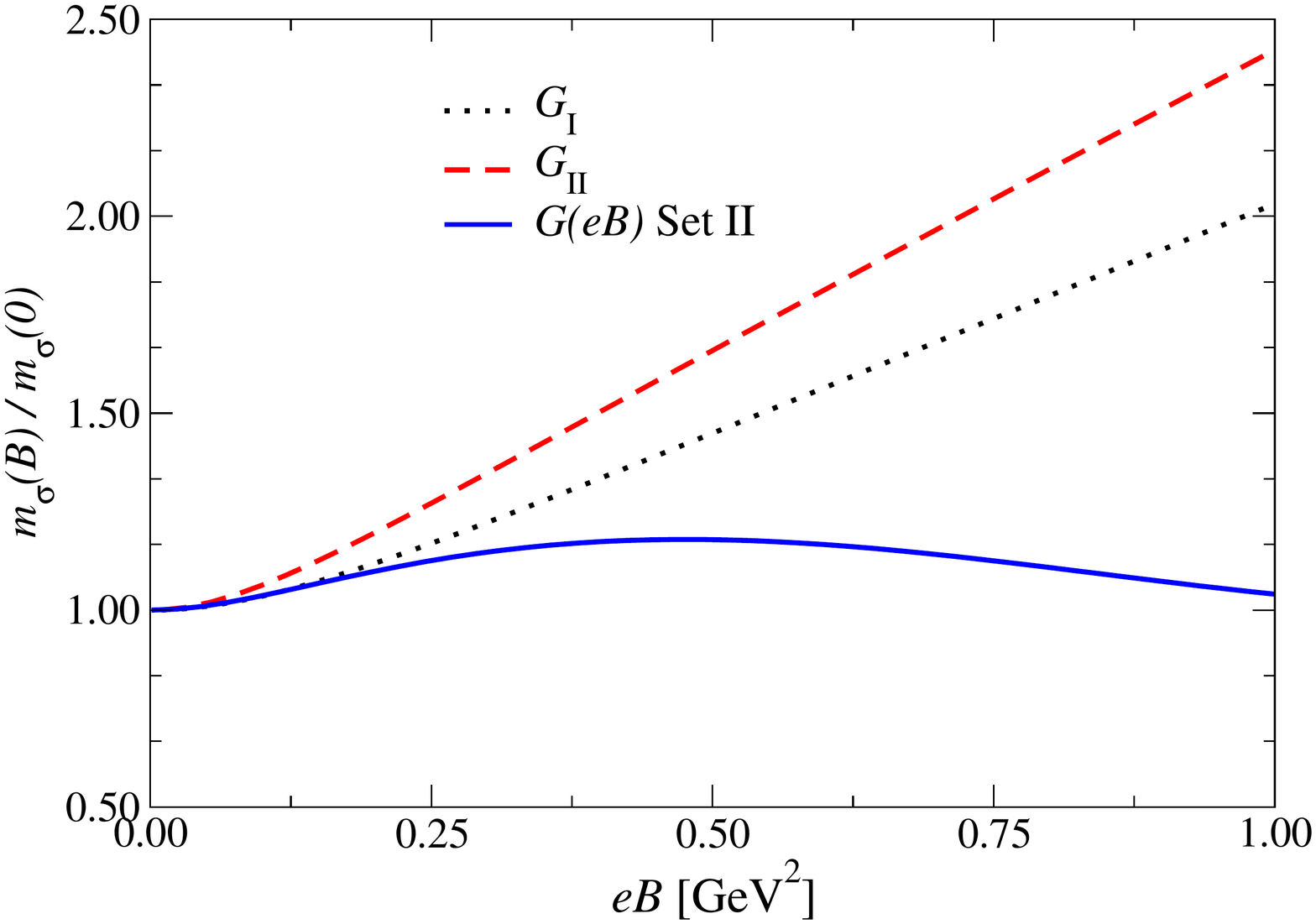}
\caption{Normalized meson masses as functions of $eB$ in the NJL model with different coupling schemes.
We also include the $m_{\pi_0}(B)$ results of \cite{simonov1}. }
\label{fig3}
\end{figure}

\subsection{ Field dependent coupling }

Let us now obtain the magnetic dependence of the NJL model coupling 
by  reproducing the lattice results of 
Ref.~\cite{fodor} for the quark condensate average at zero temperature, $ (\Sigma_u + \Sigma_d)/2$. We remark that these
precise LQCD results have been obtained for $N_f=2+1$ whereas here we are considering the two flavor case. However, in general, 
translating LQCD predictions for the $N_f=2+1$ case to $N_f=2$ effective 
models can be quite safely done because the lattice results are often divided into results for the 
light ($u$ and $d$) and strange sectors. This is particularly true in the case of the  condensates 
since only the ones related to light quarks (or rather, their average) represent the order parameter for
the chiral transition.

In  LQCD simulations, the condensates are normalized in a way which is reminiscent of Gell-Mann-Oakes-Renner 
relation (GOR), $-2m \langle \bar \psi_i \psi_i \rangle = m_\pi^2 f_\pi^2 + \dots$, so that for a given flavor 
one has
\begin{equation}
\Sigma_{i}(B) = \frac{2m}{m_\pi^2 f_\pi^2}\left[\langle \bar \psi_i\psi_i \rangle_{B} 
- \langle \bar \psi_i \psi_i \rangle_{00}\right]+1 ,
\label{sigmalatt}
 \end{equation}
with $ \langle \bar \psi_i\psi_i \rangle_{00}$ representing the quark condensate at $T=0$ 
and $B=0$. In order to fit the lattice results, the other physical quantities appearing in  
Eq.~(\ref {sigmalatt}) should be those of Ref.~\cite {fodor}; namely, $m_\pi =135~{\rm MeV}$, 
$f_\pi= 86~ {\rm MeV}$, and $m=5.5~{\rm MeV}$ so that, by invoking the GOR relation,  
one can use the LQCD value $\langle \bar \psi_i\psi_i \rangle_{00}^{1/3}=-230.55~{\rm MeV}$.

For selected values of $eB$ from zero to $1~{\rm GeV}^2$ and $T=0$, we can fit the NJL coupling to the 
corresponding values resulting from lattice QCD calculations. Then we make an interpolation 
to generate a larger set, which, in turn, is fitted to a simple shifted gaussian
for the magnetic field dependence of the coupling constant. This means a good fit to lattice simulations
for the average $(\Sigma_u + \Sigma_d)/2$ can be obtained by using 
\begin{equation}\label{eq:2}
G(eB) = \alpha ~+~ \beta ~ e^{-\gamma ~ (eB)^2}  \; \;  ,
\end{equation}
where $\alpha = 1.44373~{\rm GeV}^{-2}$, $\beta = 3.06~{\rm GeV}^{-2}$ and $\gamma = 1.31~{\rm GeV}^{-4}$. 
Note that when there is no magnetic field, $G(0) = \alpha + \beta = G_{II} = 4.50373~{\rm GeV}^{-2}$ which is
the coupling value that gives the same results as lattice QCD calculations for the condensate average at $T=B=0$.
We remark that the present ansatz is different from the one obtained in Ref. \cite {ric2}, where the fit was performed 
at the high temperatures $T>110~{\rm MeV}$. However, the interpolation procedure carried out to improve precision 
when finding the parameters for the ansatz is the same. 

\section{Numerical results}

In principle, our results are rigorously valid for $eB \le 0.4~{\rm GeV}^2$, which is the upper limit the cutoff scheme 
can account for. Hence, our results for large magnetic field strengths need to be taken as extrapolations as they give 
only a qualitative behavior in this limit.

To  carry out numerical evaluations we need the four different sets of parameters displayed in Table \ref{tab1}. Notice that sets $I$ and $II$ are used when comparing with LQCD employing physical quark masses, as in Ref. \cite {fodor}, while sets $III$ and $IV$ are more appropriate for comparisons with simulations using heavy quarks masses such as the ones performed in Refs. \cite{endrodidata1,endrodidata2}. Therefore, although the running of $G(eB)$ has been determined from a simulation with physical quark masses \cite {fodor} we can still compare with simulations which employ heavier quarks \cite{endrodidata1,endrodidata2} provided that we tune the NJL current quark masses in a appropriate way as our numerical results will demonstrate.

Note that the parameters of set I used in our calculations were determined by fitting the pion mass and its decay
constant to their empirical values $m_\pi = 138$ MeV and $f_\pi = 92.4$ MeV, respectively, and they are the
same used in the literature (see, eg, Ref. \cite {buballa}). Our set II was obtained fixing 
the NJL coupling constant that gives the same results as lattice QCD calculations for the condensate 
average at $T=B=0$. The sets III and IV were obtained just increasing the current quark masses in set I and 
II to obtain a heavy pion mass to be possible compare ours results with predictions from  recent lattice simulations. 

\begin{table}[ht]
\caption{Parameter sets for the NJL model at $T = B = 0$. The correct $eB\to0$ limit of our ansatz requires that $G_{II} = G(eB=0)$.}
\begin{center}
\begin{tabular}{ccccc}
\hline
           &  $m_{\pi_0}$ (MeV)  & $m_0$ (MeV) &   $G$ (GeV$^{-2}$) & $\Lambda$ (MeV)
\\ \hline
Set I    &    135.62          &  5.0         & $G_I$    =  4.67             &   664.3  \\
Set II   &    143.31          &  5.5         & $G_{II}$   = 4.50             &   650.0       \\
Set III  &    417               &   48.41     & $G_{III}$  = $G_{I}$             &   664.3  \\
Set IV  &    417               &  50.16     & $G_{IV} = G_{II}$             &   650.0       \\ 
\hline
\end{tabular}
\end{center}
\label{tab1}
\end{table}

\begin{figure}[h]
\centering
\includegraphics[width=8.5cm]{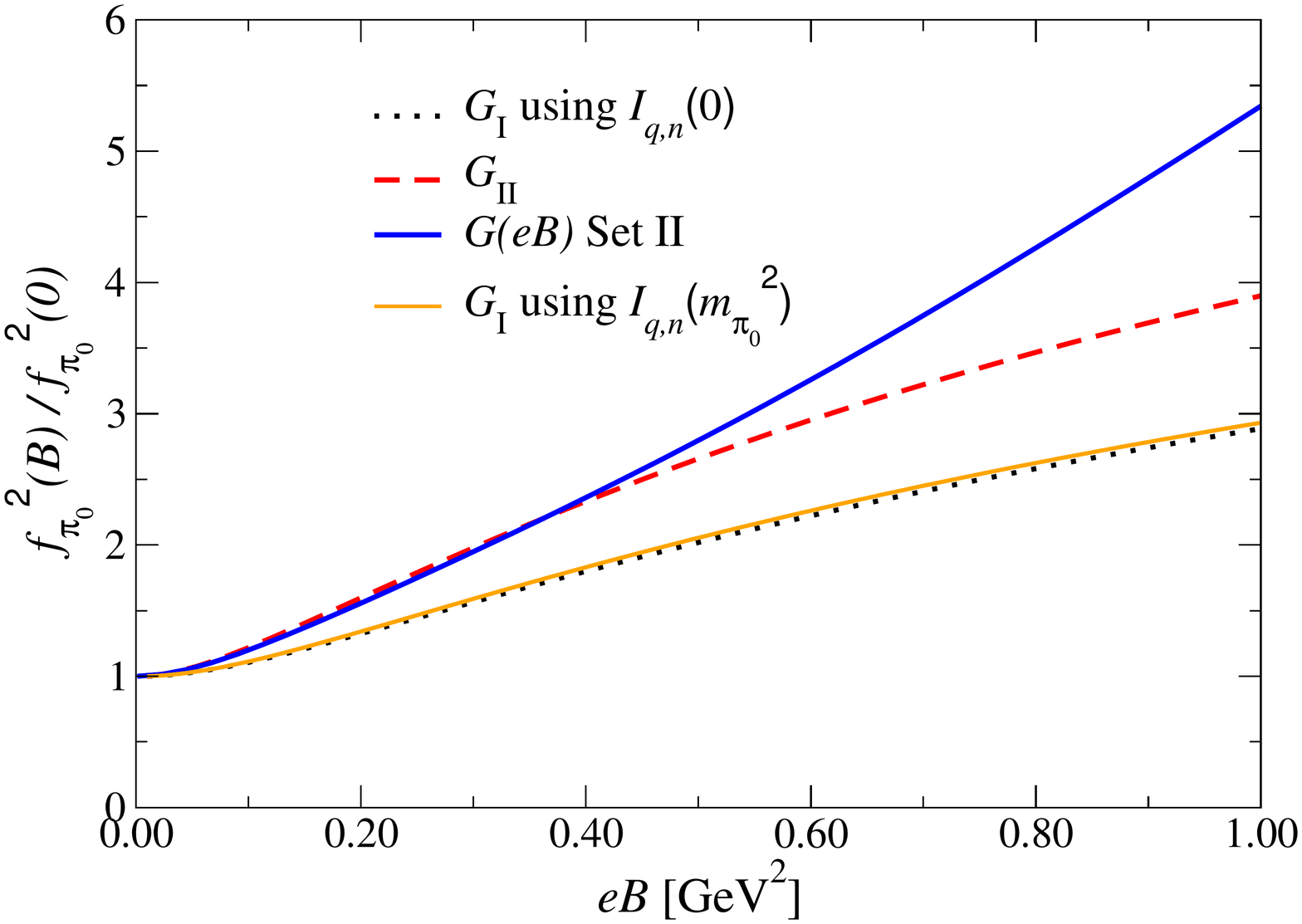}
\includegraphics[width=8.5cm]{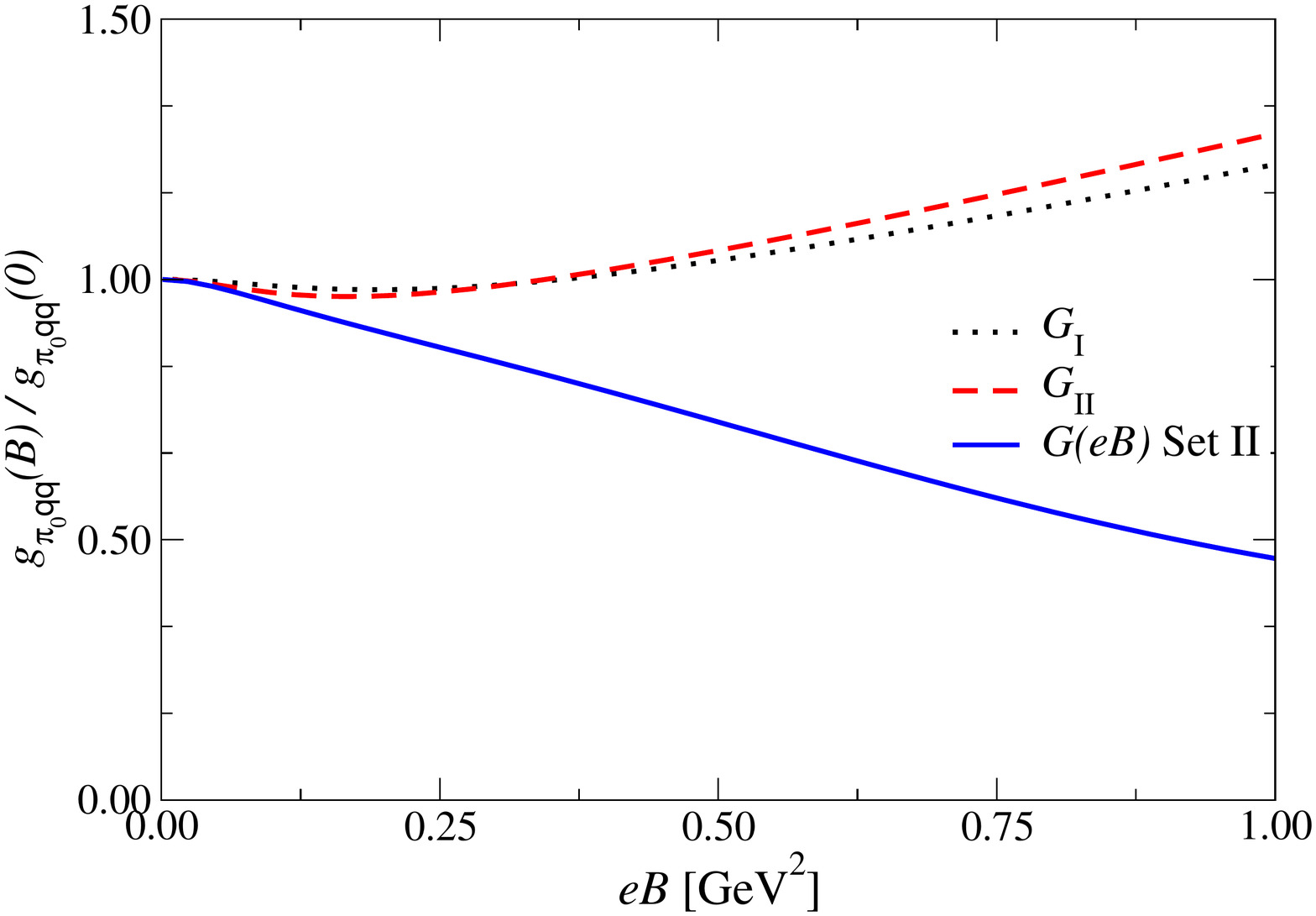}
\caption{Normalized $\pi_0$ decay constant and meson-quark coupling in the NJL model with different coupling schemes. 
For the $\pi_0$ decay constant, we also show the comparation between the RPA calculation using the complete polarization
integral as well as the approximation  } 
\label{fig4}
\end{figure}
\begin{figure}[h]
\centering
\includegraphics[width=8.5cm]{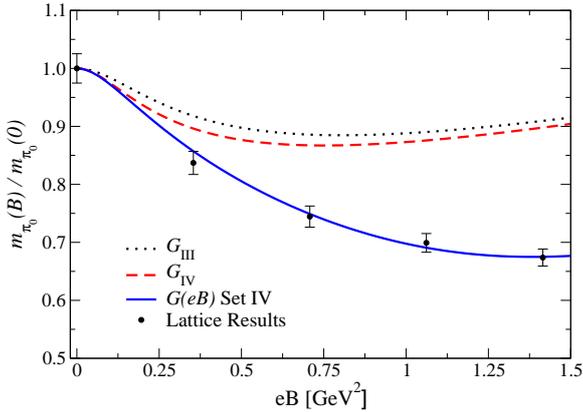}\\
\caption{Normalized neutral pion mass $m_{\pi_0}(eB)/m_{\pi_0}(0)$ in the NJL model with different coupling schemes 
and a large current quark mass compared to recent lattice results \cite{endrodidata1,endrodidata2}. }
\label{fig5}
\end{figure}

In Fig. \ref{fig1} we show  our numerical results for the average $(\Sigma_u+\Sigma_d)/2$ (upper panel)
and the difference ($\Sigma_u-\Sigma_d$) (lower panel) using the coupling constant $G_{II}$ and the
fitted coupling $G(eB)$ of eq.(\ref{eq:2}) 
in accord with the recent LQCD data \cite{fodor}. The top panel displays how the order parameter for the 
chiral transition represented by the scalar condensates increases with $B$ in a clear manifestation of the magnetic 
catalysis phenomenon. Fig. \ref{fig2} shows the  magnetized effective quark mass behavior changes drastically when 
one uses the running coupling. However, such a behavior could be anticipated by recalling that the initial 
motivation 
to adopt such coupling was to counterbalance the increase of the order parameter with $B$ so that the 
(non observable) 
effective quark mass $M \sim G  \langle {\overline \psi}_f \psi_f \rangle$ behaves differently from the
case where $G$ is fixed. 
This was particularly important at finite temperatures since in general the (pseudo)temperature is proportional to the value of the effective mass value at zero temperature (see, e.g., Ref. \cite {njlOPT})
and therefore 
IMC could be achieved by using $G(eB,T)$ in the evaluation of  $M$.

In the upper panel of Fig. \ref{fig3} we compare our results of the normalized neutral pion mass in the
MFIR scheme for different 
coupling constants $G_I$, $G_{II}$ and $G(eB)$ for $eB$ up to $1.0~{\rm GeV}^2$.  Although the curves qualitatively agree 
at very weak fields, the behavior of the neutral pion mass with $G_{I}$ and $G_{II}$ are opposite to the $G(eB)$ case 
at fields 
higher than $\approx 0.4~{\rm GeV}^2$, when the decrease of the $\pi_0$ mass is stronger in the $G(eB)$ case 
when compared 
to the  $G_{I}$ and  $G_{II}$ cases which have a slight increase. 
 We also compare our predictions for ${m_\pi}_0(B)$ with those presented in   Ref.\cite{simonov1}. 
We predict values which are about 10\% lower than those predicted in Ref. \cite{simonov1} when the eB $\lesssim$ 0.6 $GeV^2$ while 
beyond this value  our results indicate that ${m_\pi}_0(B)$ decreases in  less dramatic way.

The lower panel of Fig. \ref{fig3} shows the scalar meson mass where again the differences can be traced back 
to the fact that 
$m_\sigma \sim M$ as the figure again reveals. The results obtained with  $G(eB)$  indicate that, just like $M$, 
the sigma meson mass is quite stable (varying  less than $10 \%$ at intermediate field values)  so that the 
correlation 
length, $\xi \sim 1/m_\sigma$ also remains almost constant. On the other hand the results obtained by using
a fixed 
$G$ lead to the conclusion that the scalar mass increases so that this mode decouples while $\xi \to 0$. 

In the upper panel of Fig. \ref{fig4}, our results for the neutral pion decay constant are shown. The same 
three sets of coupling constants of Fig.\ref{fig3} have been considered. A systematic increase of $f_{\pi_0}$ as a function of 
$eB$ occurs for all 
three parameterizations and qualitatively both $G_{I}$ and $G_{II}$ constant coupling cases show a similar behavior, 
although a less dramatic increase takes place at fields greater than $0.5~{\rm GeV}^2$. Our prediction 
for $f_{\pi_0}$, 
$m_{\pi_0}$ and the quark condensates are compatible with the GOR relation. Notice  that the  validity of the approximation  $I_{q,n}(0) \approx I_{q,n}({m_{\pi_0}}^{2})$
is  confirmed since one can hardly see the difference between the calculations using $I_{q,n}(m_{\pi_0})$ or 
$I_{q,n}(m_{\pi_0}=0)$.

We have also checked 
the results for the neutral pion-quark coupling in the lower panel of Fig. \ref{fig4}, predicting a initial
decrease of its values 
up to $0.25~{\rm GeV}^2$, and then a  steadily increase with higher fields for both $G_{I}$ and $G_{II}$ cases, while for 
$G(eB)$ case we obtain a prediction of a continuous decrease which again could be anticipated by recalling 
that $g_{{\pi_0}qq} \sim M/f_{\pi_0}$ and that $f_{\pi_0}$ increases with $B$. Note also that the curve has the same 
shape as the one showed in Fig. \ref{fig2} for $M$. Finally, in Fig. \ref{fig5} we show once again our results for
the neutral 
pion mass but now, having in mind a quantitative comparison with lattice QCD results, we use the parameter 
set IV of Table I. In this parametrization the current quark mass is set equal to 50.16 MeV in order to obtain for 
$B=0$ the $\pi_0$ mass of 417 MeV, which is the value used in the lattice calculation
\cite{endrodidata1,endrodidata2}. 
Thus,  we can compare the results using different coupling constants with the recent lattice results showing that 
the behavior of the masses as a function of $eB$ is qualitatively the same as found in the top panel of
Fig. \ref{fig3}. 
That is, in accordance with LQCD predictions, our results indicate that the neutral pion remains a soft mode 
over a 
rather wide range of $B$ values. Note that Fig. \ref{fig5} indicates  that only when $G(eB)$ is  used in 
conjunction  
with a heavy current quark mass a very good quantitative agreement with recent LQCD results within the Wilson
Fermions 
Formulation \cite{endrodidata1,endrodidata2} is obtained. In those investigations, the authors discuss how the 
LQCD results 
for the pion mass in external magnetic fields depend on the critical hopping parameters, in particular, they 
show that the 
impact of their results within the Wilson Fermions Formulation has been ignored in previous works. The use 
of constant 
bare quark masses in the LQCD calculations implies that the neutral pion mass  consistently  decreases when 
$eB$ grows.
The agreement between our calculations  and the LQCD results is also a good evidence that more sophisticated 
results can be 
achieved when one assumes that the NJL SU(2) coupling constant has a dependence on $eB$ as proposed 
in Refs. \cite{ric1,ric2}.

\section{Conclusions}

The properties of magnetized neutral mesons  have been investigated using a fixed and a $B$-dependent coupling 
constant   so that model predictions and  LQCD results related to inverse 
magnetic catalysis agree. The evaluations have been performed using the two flavor NJL model following the 
RPA-MFIR framework presented in  Ref. \cite{nosso1}. One of our main  results shows  that  the $\pi_0$ remains a 
soft mode even at rather high field strengths ($\approx 1.5 \, {\rm GeV}^2$) since its mass decreases by about 
$30\%$. 
The quantitative agreement between our results and recent LQCD predictions is remarkable. Another physically 
interesting 
result refers to the behavior of the scalar meson mass which is predicted to steadily increase when  a fixed 
coupling is 
used reaching (at $eB \approx 1.0 \, {\rm GeV}^2$) a value which is two and half times higher than its value at $B=0$, 
also indicating 
a decrease of the correlation length, while our results predict that $m_\sigma$ remains quite stable. The 
different predictions 
can be easily understood by  recalling $m_\sigma \propto M \propto G \langle {\overline \psi}_f \psi_f \rangle$ 
and that, owing to 
the MC effect, the order parameter $\langle {\overline \psi}_f \psi_f \rangle$ increases within both approaches. 
On the other hand, 
the effective quark mass naturally increases when one uses a constant $G_I$ (and $G_{II}$) and remains practically stable when 
$G(eB)$ is 
considered yielding the  observed different type of behavior. 

Although  the quark mass does not necessarily represent a physical observable this is still  an interesting 
result since 
the behavior of $M$ gets directly reflected in $m_\sigma \propto 1/\xi$.  When the different model prescriptions
are 
used to  evaluate the $\pi_0$ decay the one which employs $G(eB)$ predicts an increase which is sharper than the 
one predicted by using a constant coupling value and, together with our predictions for $m_{\pi_0}$ and quark
condensates, 
observes the GOR relation. Finally, when comparing model predictions for  the meson coupling constant 
$g_{{\pi}_0 qq}$ 
we found that the use of $G(eB)$ and $G_I$ (and $G_{II}$) indicate an opposite behavior since the former predicts this quantity
to decrease 
with $B$ while the latter predicts it to increase. Once again the differences are easily understood from the 
discussions above and by recalling that this coupling is proportional to $M/f_{\pi_0}$. The results obtained
in this Letter seem to indicate that the use of a running coupling within a robust theoretical framework, 
such as the RPA-MFIR, turns the simple NJL into 
a useful tool to investigate magnetized quark matter.

\section*{Acknowledgments}
We thank B.B. Brandt  and G. Endr\"{o}di for sharing their unpublished neutral pion mass data and for 
their help in the comparison between our theoretical calculations and their LQCD results. 
R.L.S.F. acknowledges the kind 
hospitality of the Center for Nuclear Research at Kent State University, where part of this work has been done.
This work was supported by CNPq; grants No. 475110/2013-7 (RLSF), 232766/2014-2 (RLSF), 308828/2013-5 (RLSF), 
306195/2015-1 (VST), 307458/2013-0 (SSA), 303592/2013-3 (MBP); FAPESP, grant No. 2014/04975-9 (VST); and CAPES.

\end{document}